\documentclass[acmsmall]{acmart}
\usepackage{graphicx}
\AtBeginDocument{%
  }

\begin{document}

\setcopyright{cc}
\setcctype{by}
\acmJournal{PACMCGIT}
\acmYear{2026} \acmVolume{9} \acmNumber{3} \acmArticle{43}
\acmMonth{7} \acmDOI{10.1145/3816080}

\title{Electrospun Fields: 3D Nano-Fiber Material Computation as Design Method}

\author{Wai Lok Wan}
\orcid{0009-0007-8391-0841}
\email{justinjw@mit.edu}
\affiliation{%
  \institution{MIT}
  \city{Cambridge}
  \state{MA}
  \country{USA}
}

\author{Ayah Mahmoud}
\orcid{0009-0000-2273-285X}
\email{ayahm@mit.edu}
\affiliation{%
  \institution{MIT}
  \city{Cambridge}
  \state{MA}
  \country{USA}
}

\author{Sergio Mutis}
\orcid{0009-0006-4074-0734}
\email{smutis@mit.edu}
\affiliation{%
  \institution{MIT}
  \city{Cambridge}
  \state{MA}
  \country{USA}
}

\author{Avantika Velho}
\orcid{0009-0003-5253-7467}
\email{avantikavelho1@gmail.com}
\affiliation{%
  \institution{Harvard University}
  \city{Cambridge}
  \state{MA}
  \country{USA}
}

\author{Annie Xing}
\orcid{0009-0003-3743-5876}
\email{anniexin@mit.edu}
\affiliation{%
  \institution{Harvard University}
  \city{Cambridge}
  \state{MA}
  \country{USA}
}

\author{Behnaz Farahi}
\authornote{Director of the Critical Matter Group: \url{https://www.media.mit.edu/groups/critical-matter/overview/}}
\orcid{0000-0002-5672-6949}
\email{behnaz_f@media.mit.edu}
\affiliation{%
  \institution{MIT Media Lab}
  \city{Cambridge}
  \state{MA}
  \country{USA}
}
\renewcommand{\shortauthors}{Wan et al.}

\begin{abstract}

\textit{Electrospun Fields} presents a material–computational approach to design in which form emerges through the interaction of matter, electric fields, and computation rather than through explicit geometric prescription. Departing from geometry-driven fabrication paradigms, this work explores robotic electrospinning of biodegradable nanofibers onto three-dimensional conductive structures, treating electrospinning as a spatial, field-based modeling process. Bio-compatible materials—including keratin, silk, and biodegradable synthetic polymers—are investigated to establish an operating envelope for material-driven form generation.

Through systematic studies of scaffold geometry, we show that topological curvature plays a critical role in material accumulation: while convex geometries support consistent deposition, concave regions exhibit field shielding that prevents fibers from penetrating geometric valleys. This limitation directly motivates the development of a custom robotic electrospinning platform capable of dynamically reorienting the emitter to access complex three-dimensional topologies.

In this framework, conductive scaffolds condition electric field distributions rather than define form directly. Electric fields operate as a physical computation layer, and electrospun fibers act as a material rendering of force interactions. Properties such as fiber density, porosity, orientation, and thickness emerge through the coupled dynamics of field conditions, scaffold geometry, and material behavior. The resulting structures are grown rather than assembled, foregrounding emergence, material intelligence, and field-mediated form generation, and offering an alternative to geometry-centric fabrication in art and design.

\end{abstract}

\ccsdesc[500]{Applied computing~Fine arts}
\ccsdesc[300]{Computer systems organization~Robotics}
\ccsdesc[100]{Computing methodologies~Modeling and simulation}
\ccsdesc[300]{Computer systems organization~Robotic components}

\keywords{material computation, robotic fabrication, electrospinning, field-conditioned deposition, conductive scaffolds, biodegradable polymers, nanofiber membranes, form-finding}

\begin{teaserfigure}
  \includegraphics[width=\textwidth]{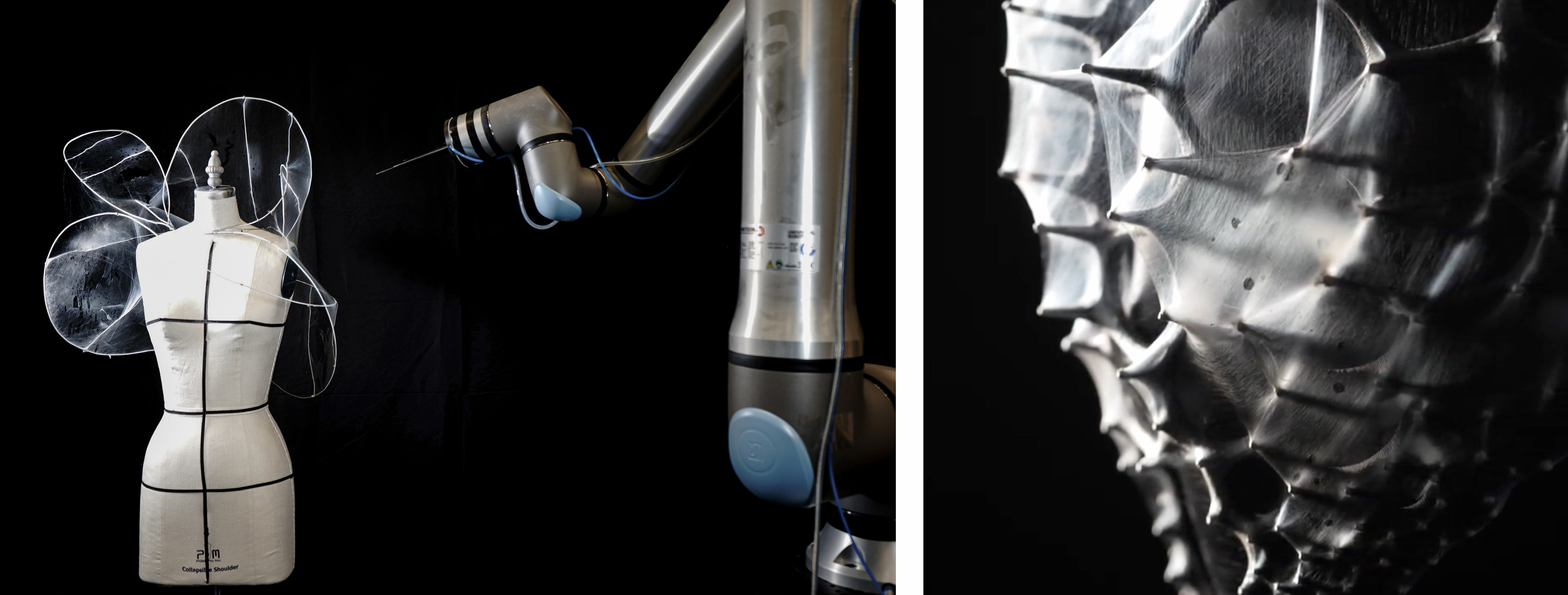}
  \caption{(Left) UR20 Robotic arm electrospinning into sculptural garment on wire. (Right) close-up of electrospun conductive PLA mask.}
  \label{fig:teaser}
\end{teaserfigure}

\maketitle

\section{Introduction}

Contemporary computational fabrication is largely shaped by geometry-driven, top-down paradigms in which digital models prescribe form and materials are treated as passive substrates optimized to comply with predefined specifications. In the translation from material to model, many material properties are abstracted or lost within computational representation. While these approaches enable high levels of precision and repeatability, they constrain material behavior rather than engaging it, limiting opportunities for emergence as part of the design process.

Material computation offers an alternative framework in which matter is understood as an active participant in form-making. Rather than privileging geometric control, it emphasizes interactions between material properties, environmental forces, and designed constraints. Form, in this context, is not imposed but emerges through dynamic processes, reframing fabrication as a relational and process-driven practice.

Electrospinning provides a fertile ground for exploring material computation. The technique uses electric fields to draw nano fibers from polymer solutions, producing fibrous structures whose behavior is highly sensitive to voltage, distance, geometry, gravity, and material viscosity. These sensitivities allow electrospinning to function not merely as a deposition technique, but as a spatially responsive material system.

However, existing electrospinning research has largely prioritized uniformity, precision, and two-dimensional control, focusing on flat substrates and tightly regulated outcomes for biomedical and industrial applications. As a result, the spatial and expressive potentials of electrospinning remain underexplored. By re-situating electrospinning within a material–computational framework, this paper challenges the reduction of electrospinning to controlled deposition and introduces a robotic method for three-dimensional form generation driven by electric fields and material behavior.

We contribute (1) a reusable design method for field-conditioned electrospinning, (2) a bio-compatible material catalog defining an operating envelope, (3) a scaffold taxonomy linking geometry to deposition behavior, and (4) a reproducible robotic platform enabling non-planar deposition on concave topologies.

\section{State of the Art}
\subsection{Material Computation}

Over the past two decades, computational design and fabrication research has increasingly moved beyond purely geometry-driven workflows toward approaches that integrate material behavior as an active driver of form.

Work by Neri Oxman articulated material computation as a design framework, enabling form to emerge through the interaction of material properties and environmental forces \citep{oxman2007materialbaseddesigncomputation,oxman2009materialecology,oxman2010structuringmateriality}. In parallel, Achim Menges framed material computation as a morphogenetic design paradigm in which material behavior and computational processes co-develop form \citep{menges2012materialcomputation}, as illustrated by the ICD/ITKE Research Pavilions since the 2010s \citep{icditkeResearchPavilions}. In generative art, Lomas explored emergence through simulated cellular growth, producing complex organic sculptures from minimal rules governing cell forces and nutrient accumulation—foregrounding morphogenesis and self-organization as artistic methods in which form is grown rather than designed \citep{lomas2014cellularforms}.

Within SIGGRAPH art works, this reframing of matter as an active participant has a lineage spanning over a decade. Beesley's \textit{Hylozoic Soil} \citep{beesley2009hylozoicsoil}—a room-scale responsive sculpture whose digitally fabricated meshwork breathed and pulsed through shape-memory alloy actuators—demonstrated as early as 2009 that material systems can exhibit agency when coupled with environmental sensing. Subsequent work sharpened this premise into fabrication-specific inquiry: Klein \citep{klein2018augmentedfauna} staged a deliberate dialogue between molten glass and 3D-printed form, showing that material behavior can inflect and redirect digital intent; Starrett et al. \citep{starrett2018datamaterialization} employed electroforming—a field-mediated electrochemical deposition process—to let electric current, rather than toolpath, drive material accumulation; and Harvey et al.'s \textit{Weaving Objects} \citep{harvey2019weavingobjects} demonstrated that computationally programmed yarn can co-determine three-dimensional textile form through its own tension and settlement behavior.

More recent SIGGRAPH art works push further, progressively ceding formal control to material and environmental processes. Lee and Llach \citep{lee2020hybridembroidery} introduced vision-guided embroidery that adapts to material feedback in real time, closing the loop between stitch and sensor. Montero and De~Berduccy \citep{montero2021spinningconductors} reframed the Andean loom as an indigenous computational system, revealing that textile fabrication encodes computational logic well beyond the digital. Meiklejohn et al.'s self-shaping textiles \citep{meiklejohn2022wovenbehavior} compute their 3D form through thermal response alone—the designer programs yarn, not shape. DiBlasi et al.'s \textit{Beauty} \citep{diblasi2023beauty} extended this principle to biological media, treating bacterial growth as a material-computational agent that drives audio-visual composition. And Elran et al. \citep{elran2023finemotorskills} framed the robotic fabricator itself as a generative co-author, whose collaboration with clay plasticity yields structures that neither party prescribed.

Together, these works mark a shift in SIGGRAPH art, and material computation discourse broadly, from enforcing geometry to conditioning material behavior. Yet the material systems explored—molten glass, woven yarn, embroidery thread, bacteria, wet clay—remain mediated by mechanical contact, thermal stimulus, or direct toolpath, leaving electrostatic field–based fabrication, in which invisible force fields drive material deposition, comparatively unexplored.

\subsection{Electrospinning}

Electrospinning is a fiber fabrication technique in which electric fields draw continuous jets from polymer solutions to produce nonwoven fibers with diameters ranging from the micro- to nanoscale. Since early demonstrations of stable polymer nanofibers \citep{doshi1995electrospinning,reneker1996nanometre}, electrospinning has been widely adopted for surface-area applications including filtration, tissue engineering, and surface coatings \citep{li2004reinventing,xue2019electrospinning}.

Across these domains, electrospinning research has largely prioritized uniformity, repeatability, and performance, relying on static planar collectors and tightly controlled parameters to produce homogeneous 2D fiber mats. While subsequent work has explored material innovation and deployment—such as biomaterial systems including silk fibroin, keratin, and collagen \citep{persano2013upscaling}, portable and in-situ devices for deposition onto skin \citep{xu2019portableelectrospinning}, aligned fiber arrays using split collectors \citep{li2003alignedarrays}, and hybrid electrospinning–additive manufacturing systems \citep{rivera2019desktopelectrospinning}—most approaches largely minimize geometric and field variability.

As a result, electrospinning has rarely been investigated as a three-dimensional, geometry-conditioned fabrication process. Variations in field geometry, collector configuration, and spatial deposition are typically suppressed rather than leveraged, leaving a gap in its use within material computation, where fields, geometry, and material behavior could function as coupled design variables rather than sources of noise.

\section{Design Method: \textit{Electrospun Fields}}

Against this backdrop, \textit{Electrospun Fields} proposes electrospinning as a material-computational design method. The method treats conductive geometries as computational boundary conditions that modulate electric fields, enabling fiber density, orientation, porosity, and thickness to emerge as a material rendering of voltage-material-geometry interaction. In this way, electrospinning is reframed as a three-dimensional process of computing matter rather than depositing material onto predefined forms.

\textit{Electrospun Fields} presents a reusable design methodology composed of three components: (1) an electrospinning platform, either conventional or robotic, that defines the spatial degrees of freedom of deposition; (2) a taxonomy of bio-compatible electrospinnable polymers that establishes the operating envelope of the system; and (3) a taxonomy of scaffold geometries and associated field-conditioned material behaviors, enabling designers to anticipate how scaffold form influences electrospun 3D membrane formation.

\section{Electrospinning Platforms}

\textit{Electrospun Fields} is implemented through two electrospinning hardware configurations: a conventional fixed electrospinning setup and a custom, DIY robotic electrospinning platform that extends spatial reach, control, and field-conditioning capabilities. Designers may employ either configuration depending on the desired scale, geometric complexity, and degree of spatial control.

\subsection{Fixed Electrospinning Setup}

\begin{figure} [h]
  \includegraphics[width=0.85\textwidth]{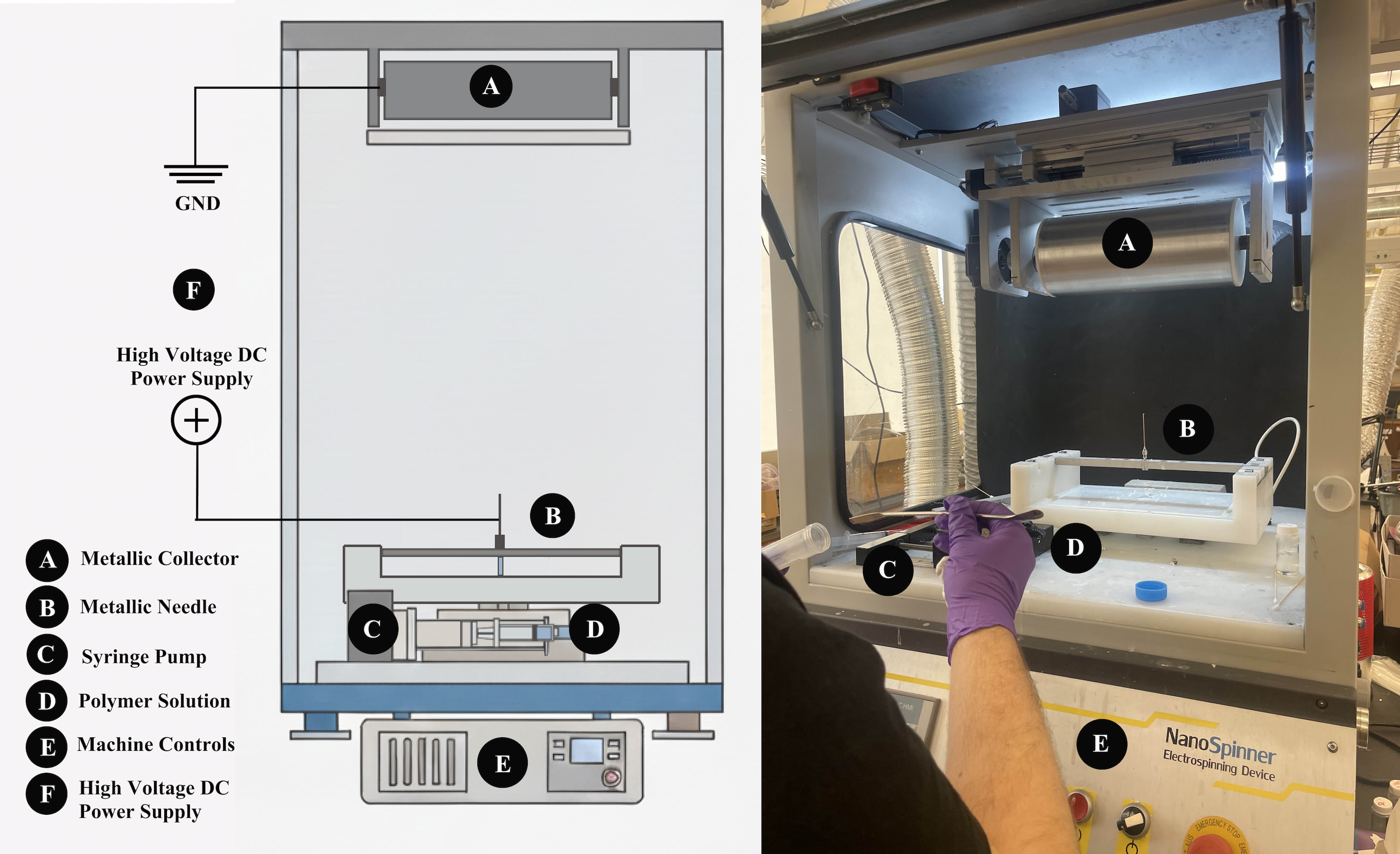}
  \caption{Traditional Electrospinning Setup}
  \label{fig:2_electrospinning-fixed}
\end{figure}

The fixed electrospinning setup follows a conventional laboratory configuration and consists of a syringe pump delivering polymer solution through a metallic needle connected to a high-voltage power supply, with fibers collected on a grounded conductive substrate at a fixed working distance (Figure~\ref{fig:2_electrospinning-fixed}). Flow rate, applied voltage, and needle-collector distance define range, jet stability, and fiber morphology. Its static geometry and high repeatability make it well suited for isolating material and field variables, while its planar configuration limits volumetric deposition and large-scale spatial variation.

\subsection{Custom UR20 Electrospinning Platform}

\begin{figure} [H]
  \includegraphics[width=\textwidth]{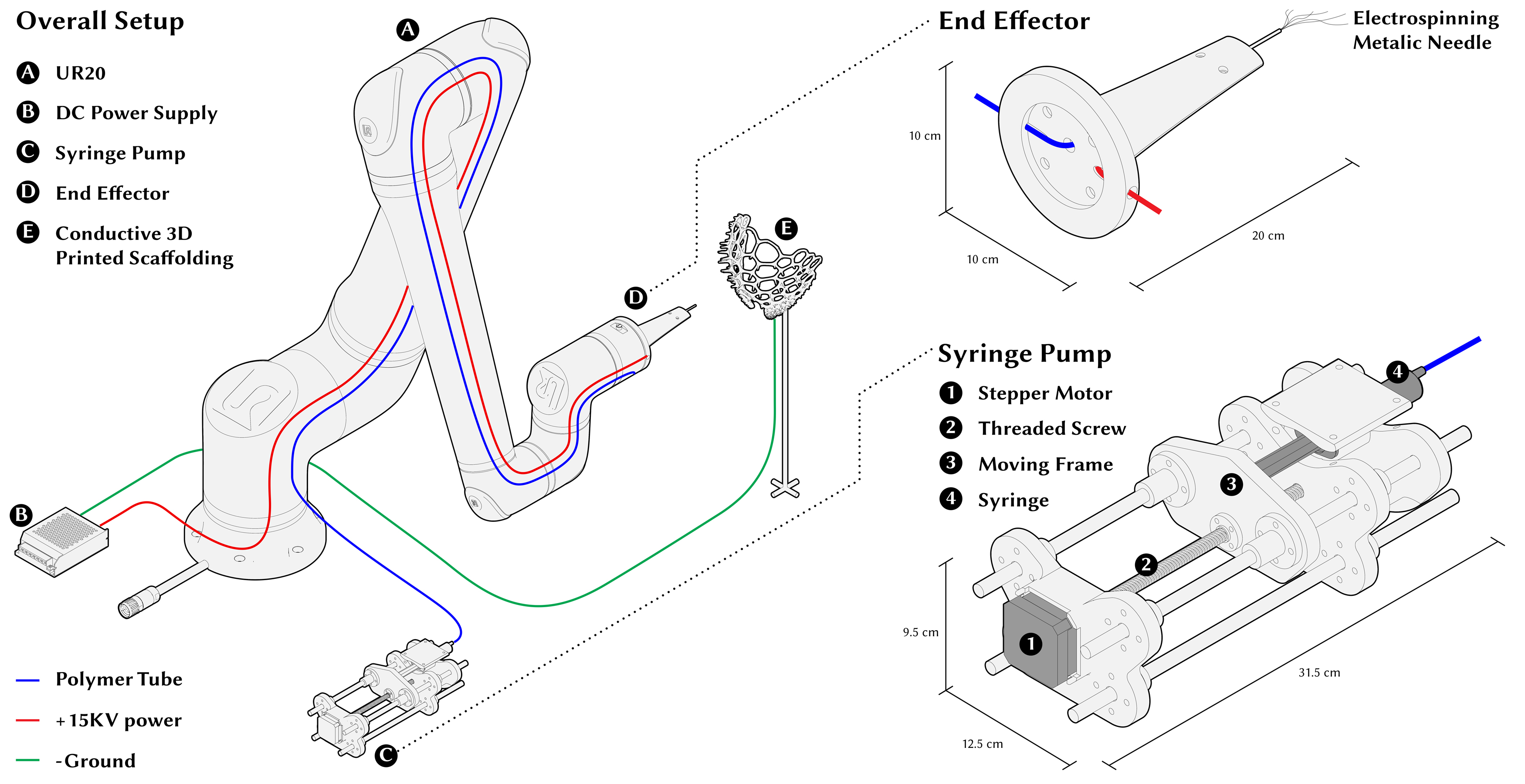}
  \caption{Custom UR20 Electrospinning Platform}
  \label{fig:3_electrospinning-UR20}
\end{figure}

To enable three-dimensional, geometry-conditioned electrospinning beyond the constraints of fixed setups, we contribute a custom electrospinning end-effector integrated with a six-axis Universal Robots UR20 robotic arm (Figure~\ref{fig:3_electrospinning-UR20}). The end-effector incorporates a localized syringe pump driven by a high-torque stepper motor, ensuring consistent polymer flow independent of tool orientation, and routes a high-voltage connection directly to the robot-mounted needle, allowing it to function as a mobile emitter relative to grounded conductive geometries. 

This platform decouples electrospinning from a fixed vertical axis and introduces full kinematic control over emitter position, orientation, working distance, and traversal velocity. By synchronizing robot motion with electrospinning parameters, the electric field becomes spatially programmable: conductive scaffold geometry, deposition angle, and distance actively condition fiber accumulation, orientation, and density. The robotic platform underpins the larger-scale, non-planar artifacts presented in Section 7, extending electrospinning from planar deposition toward choreographed, three-dimensional field-conditioned fabrication. Detailed assembly instructions and toolpath-generation code are documented in an open-source repository available at \url{https://github.com/Critical-Matter-MIT-Media-Lab/Custom-3D-Electrospinning}. 

\textit{Safety Note: As this process utilizes high-voltage direct current (up to 25 kV), proper electrical insulation and grounding are strictly required. To prevent electric shock, operators must maintain a safe standoff distance and avoid any physical contact with the emitter, polymer jet, or scaffold while the system is energized.}
\section{Bio-Compatible Electrospinning Material Catalog}
Under this method, polymer selection constitutes a primary design decision. Because electrospinning behavior is highly sensitive to solution chemistry, conductivity, and viscosity, small variations in material formulation can produce qualitatively different outcomes.

We focus on bio-compatible polymers and biomaterials to align the method with material systems that can support future applications in wearable interfaces, soft structures, and sustainable fabrication. Four material systems were explored: polyethylene oxide (PEO), polyvinyl alcohol (PVA), keratin-based blends, and silk-based blends.

\subsection{Solution Preparation}

\begin{figure}[h]
  \includegraphics[width=\textwidth]{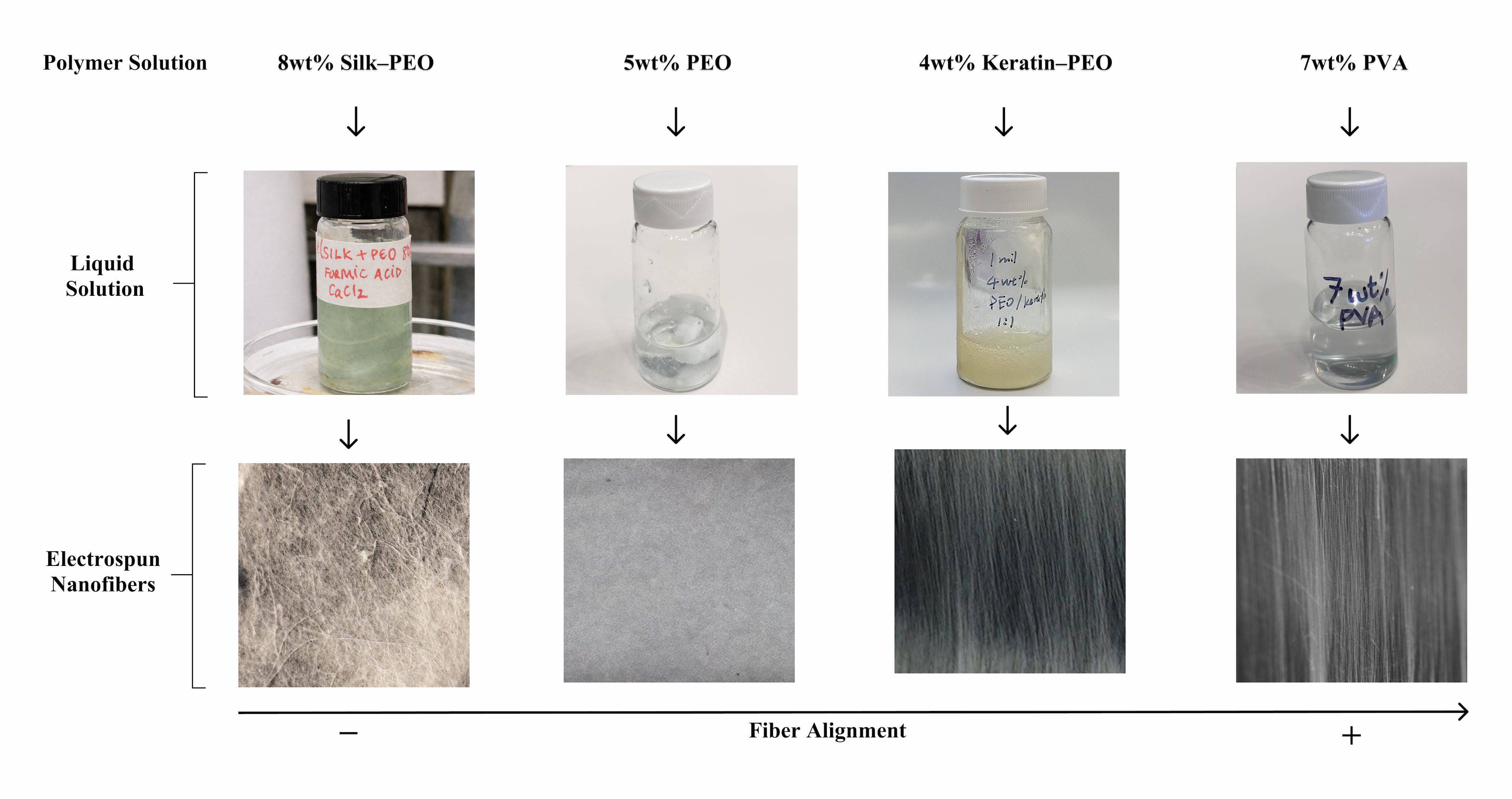}
  \caption{Material Catalogue: polymer solutions \& resultant electrospun nanofibers.}
  \label{fig:4_solution-prep}
\end{figure}

Synthetic polymer solutions of PEO and PVA were prepared as baseline materials due to their well-documented electrospinnability. PEO powder (100{,}000 g/mol) was dissolved in distilled water. Solutions ranging from 2--6 wt\% were tested; 5 wt\% yielded the most stable and consistent fiber formation. PVA solutions were prepared using an identical process, with concentrations between 5--7 wt\% evaluated. The 7 wt\% PVA solution produced the most robust fibers.

For keratin-based electrospinning, initial mixtures of dissolved human hair on lithium bromide  produced unstable jets (electrospraying) rather than continuous fiber formation. To stabilize the process, hydrolyzed keratin was blended with PEO as a carrier polymer, yielding a 4 wt\% keratin--PEO solution.

Silk solutions were prepared by dissolving 0.78 g of degummed silk fibers and 0.2 g of PEO in 9 mL of 99\% formic acid containing 0.24 g anhydrous calcium chloride, yielding an 8 wt\% polymer solution. PEO again served as a stabilizing agent to the protein solution.

\subsection{Electrospinning Tests and Qualitative Evaluation}

All material systems were electrospun onto 40$\times$40mm 3D-printed conductive scaffolds with the fixed electrospinning setup. PEO, PVA, and keratin--PEO solutions were spun at 15 kV with a flow rate of 0.25 mL/hr, while silk--PEO solutions were spun at 25 kV and 0.15 mL/hr.

PEO (5 wt\%) produced consistent, fine fibers that accumulated into visually homogeneous membranes, with fiber directionality observable under the microscope. The keratin--PEO (4 wt\%) blend generated thicker fibers with stronger directional alignment and pronounced optical iridescence; however, these membranes exhibited reduced mechanical strength and fractured over time.

PVA (7 wt\%) produced the most mechanically robust fibers, with large, highly aligned strands visible to the naked eye. Deposition occurred rapidly, producing visible accumulation within five minutes. Extended spinning increased membrane thickness and strength but reduced visual fiber differentiation, resulting in dense, removable sheets.

Silk--PEO electrospinning produced long fibers with significant diameter variation and limited directional consistency. Fibers frequently entangled in flight before deposition, leading to locally disordered accumulation. Despite this variability, the resulting membranes conformed closely to the minimal surface geometries implied by the conductive scaffolds, suggesting strong coupling between field geometry and material behavior even under unstable jet conditions.

\begin{table}[t]
\centering
\small
\caption{Summary of electrospinning mixtures and observed deposition behavior across four materials (PEO, PVA, Keratin--PEO, and Silk--PEO), including deposition speed, stability, fiber size, fiber alignment, and durability.}
\setlength{\tabcolsep}{2pt} 
\renewcommand{\arraystretch}{1.25} 

\begin{tabular}{|l|p{3.5cm}|p{1.6cm}|c|p{1.2cm}|p{1.6cm}|c|}
\hline
\textbf{Material} &
\textbf{Mixture (per 10ml)} &
\textbf{Deposition Speed} &
\textbf{Stability} &
\textbf{Fiber size} &
\textbf{Fiber alignment} &
\textbf{Durability} \\
\hline
PEO (5 wt\%) &
0.5 g PEO + 9.5 mL distilled water &
Medium &
Stable &
uniform, very fine &
Medium &
Medium \\
\hline
PVA (7 wt\%) &
0.7 g PVA + 9.3 mL distilled water &
High &
Stable &
uniform, fine &
High &
High \\
\hline
Keratin--PEO (4 wt\%) &
0.32 g keratin + 0.08 g PEO + 9.6 mL distilled water &
Medium &
Stable &
uniform, fine &
High &
Low \\
\hline
Silk--PEO (8 wt\%) &
0.78 g silk fibers + 0.20 g PEO + 0.24 g CaCl$_2$ + 9.0 mL formic acid &
Low &
Unstable &
varied, fine to coarse &
Low &
High \\
\hline
\end{tabular}
\label{tab:material-mixtures}
\end{table}

Table~\ref{tab:material-mixtures} summarizes the material systems explored, their compositions, electrospinning parameters, and observed behaviors. This table functions as a material reference catalog, which was applied to the geometry-conditioned deposition behaviors studied in Section~6.

\section{Field-Conditioned 3D Material Deposition}

\begin{figure} [h]
  \includegraphics[width=1\textwidth]{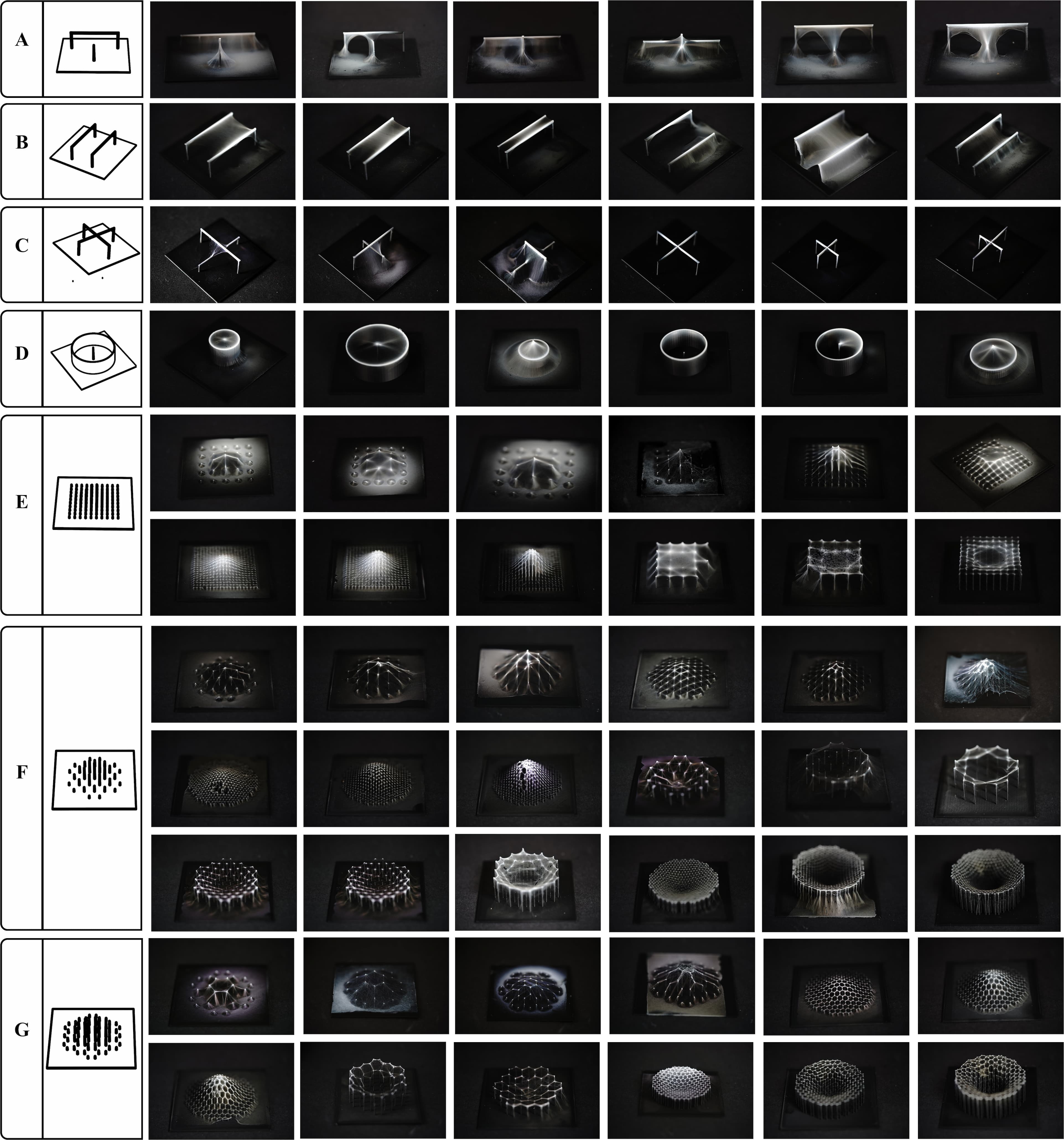}
  \caption{Taxonomy catalogue of 3D printed scaffolds and electrospun prototypes. Taxonomy Groups: (A) Line and Point (B) Parallel Lines (C ) Perpendicular Lines (D) Circles and Point (E) Square Grid (F) Triangular Grid (G) Hexagonal Grid}
  \label{fig:5_catalog}
\end{figure}

To decode the morphological language of 3D electrospinning, we electrospun onto a taxonomy of conductive substrates (Figure~\ref{fig:5_catalog}), testing how fundamental geometric operations modulate the electric field and resulting fiber deposition. The taxonomy progresses deliberately from primitives to lattice systems, isolating fundamental field--deposition relationships before combining them in the demonstrative artifacts of Section~7, where scaffolds incorporate multiple computational geometric operations across complex three-dimensional topologies.

All scaffolds were 3D-printed in carbon-black-doped conductive PLA, providing both mechanical support and a grounded collector during electrospinning.

\subsection{Computational Scaffolding Design}

We first analyzed geometric primitives---isolated points, single curves, and perpendicular intersections. In linear configurations, we varied inter-element spacing and line length to identify the proximity threshold at which distinct accumulations merge into a continuous membrane. For circular geometries, we tested radii from 1--3\,cm, examining how curvature tightness shifts deposition from tangential alignment to cross-void bridging.

We then evaluated hybrid compositions that combine disparate geometric types. Pairing point singularities with linear edges, we varied Euclidean distance from 1--3\,cm to characterize field interference and point-attractor effects along adjacent paths. Vertical differentiation was introduced by extruding nodes into spikes with a $\pm$1\,cm height differential relative to the base geometry.

Finally, we tested grid-based lattices across square, triangular, and hexagonal topologies. For each, we modulated cell density with apertures from 0.25--1\,cm, identifying the maximum span fibers could bridge before failure and whether deposition forms volumetric catenary structures versus surface coating. We also evaluated substrates with topological curvature, including deep concavities and sharp convexities.

\subsubsection{Taxonomy of Deposition Behavior}

To formalize the relationship between scaffold design and material expression, we operationalize Oxman’s material computation descriptors \citep{oxman2010materialbaseddesigncomputation_thesis} as an observational framework based on three key material–computational characteristics of electrospun textiles. (1) First, \textit{anisotropy} describes the directionality of fiber alignment, ranging from strong, visible parallelism determined by field lines to weak, random, felt-like accumulation. (2) Second, \textit{heterogeneity} defines the spatial variance of fiber density, where strong heterogeneity produces gradients of opacity and thickness, while weak heterogeneity implies uniform coverage. (3) Finally, \textit{hierarchical structuring} describes the emergence of secondary structures, such as distinct surface patterns or multi-level layering, independent of the base membrane.

\begin{figure} [h]
  \includegraphics[width=0.85\textwidth]{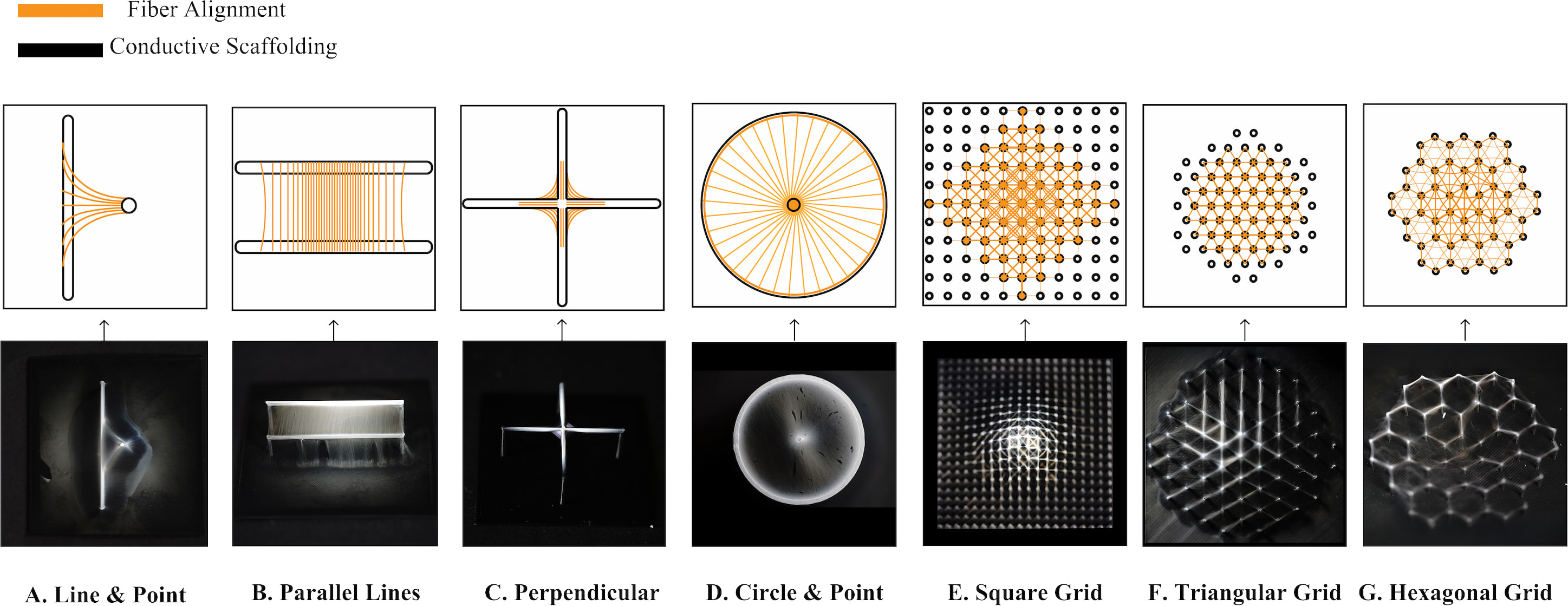}
  \caption{Fiber alignment diagrams by Taxonomy Group, highlighting geometry-conditioned, emergent directionality patterns}
  \label{fig:6_fiber-alignement}
\end{figure}

\begin{figure} [h]
  \includegraphics[width=0.85\textwidth]{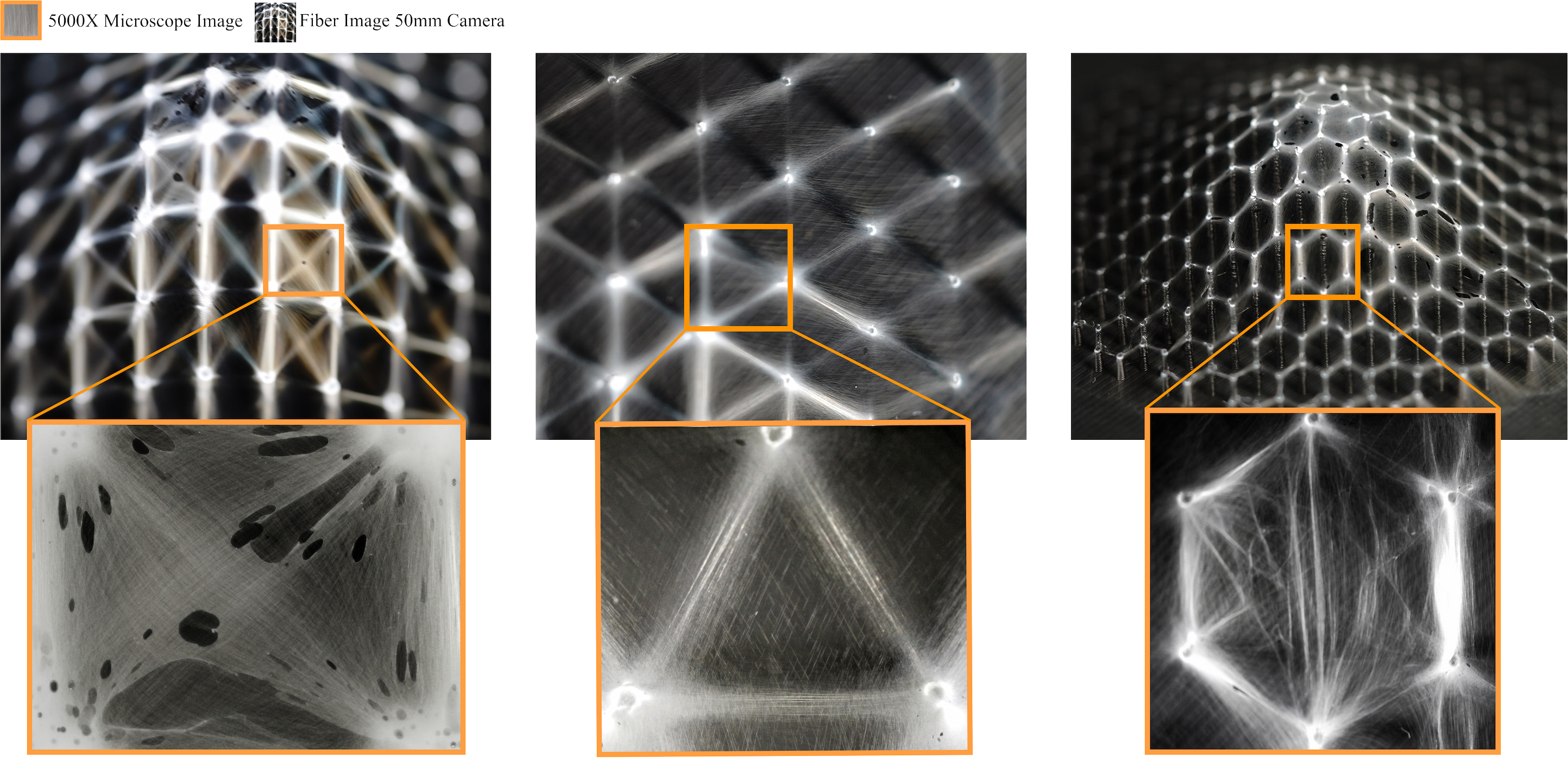}
  \caption{Close up images of electrospun membrane on (from left to right) square grid, triangular grid, and hexagonal grid scaffolds}
  \label{fig:7_Close-Ups}
\end{figure}

Observations of geometric primitives revealed that isolated points function as primary field attractors. These configurations yielded radial gradients, characterized by fibers extending outward from the singularity to form conical, tent-like membranes. Here, pronounced heterogeneity emerged as fiber density diminished with distance from the attractor, directly reflecting the rapid decay of field intensity.

Scaffolds composed of parallel bars consistently produced linear anisotropy, resulting in continuous single membranes with strong uniform alignment. However, a distinct phenomenon emerged within perpendicular curve configurations: rather than a simple superposition of linear accumulations, fibers self-organized into continuous hyperbolic minimal surfaces suspended between orthogonal axes. This suggests that the interaction between electrostatic forces and fiber tension effectively solves for equilibrium states, mirroring soap-film mechanics.

The most complex behavior was observed in grid-based lattice systems. These structures demonstrated the strongest combination of anisotropy and hierarchical structuring. Crucially, the integration of vertical nodes (spikes) transformed the deposition plane into a volumetric field; fibers were suspended in tension, bridging spike tips to form catenary structures. This created a dual-system morphology: dense, opaque structural "ridges" formed along high-field conductive paths, while thin, translucent membranes suspended themselves in low-field voids.

Finally, we analyzed the impact of topological curvature. Convex scaffold geometries facilitated consistent fiber accumulation, as diverging electric field lines exposed a greater surface area to deposition. In contrast, concave geometries exhibited pronounced field shielding: fibers failed to penetrate deep recesses and instead bridged across elevated features, forming a cap-like membrane over the void. This limitation, where static emitters cannot deposit material into geometric valleys, directly motivated the development of the robotic DIY electrospinner described in Section 4, which enables dynamic emitter reorientation and access to concave topologies inaccessible to fixed-axis systems.

Taken together, these results demonstrate a visual input–output system: the scaffold conditions the field, and the material renders the forces, shifting the designer's role from prescribing geometry to programming the conditions under which form emerges.

\section{Applications \& Demonstrative Artifacts}

To validate the \textit{Electrospun Fields} methodology, we developed three demonstrative artefacts that translate the findings of Section 6 into applied design contexts. Progressing from intimate body-interface to spatial volume and finally to dynamic display, these projects test the scalability of material computation and explore its potential as a medium for artistic practice.

\subsection{\textit{Second Selves}: Electrospun Keratin Mask}

\begin{figure} [h]
  \includegraphics[width=0.85\textwidth]{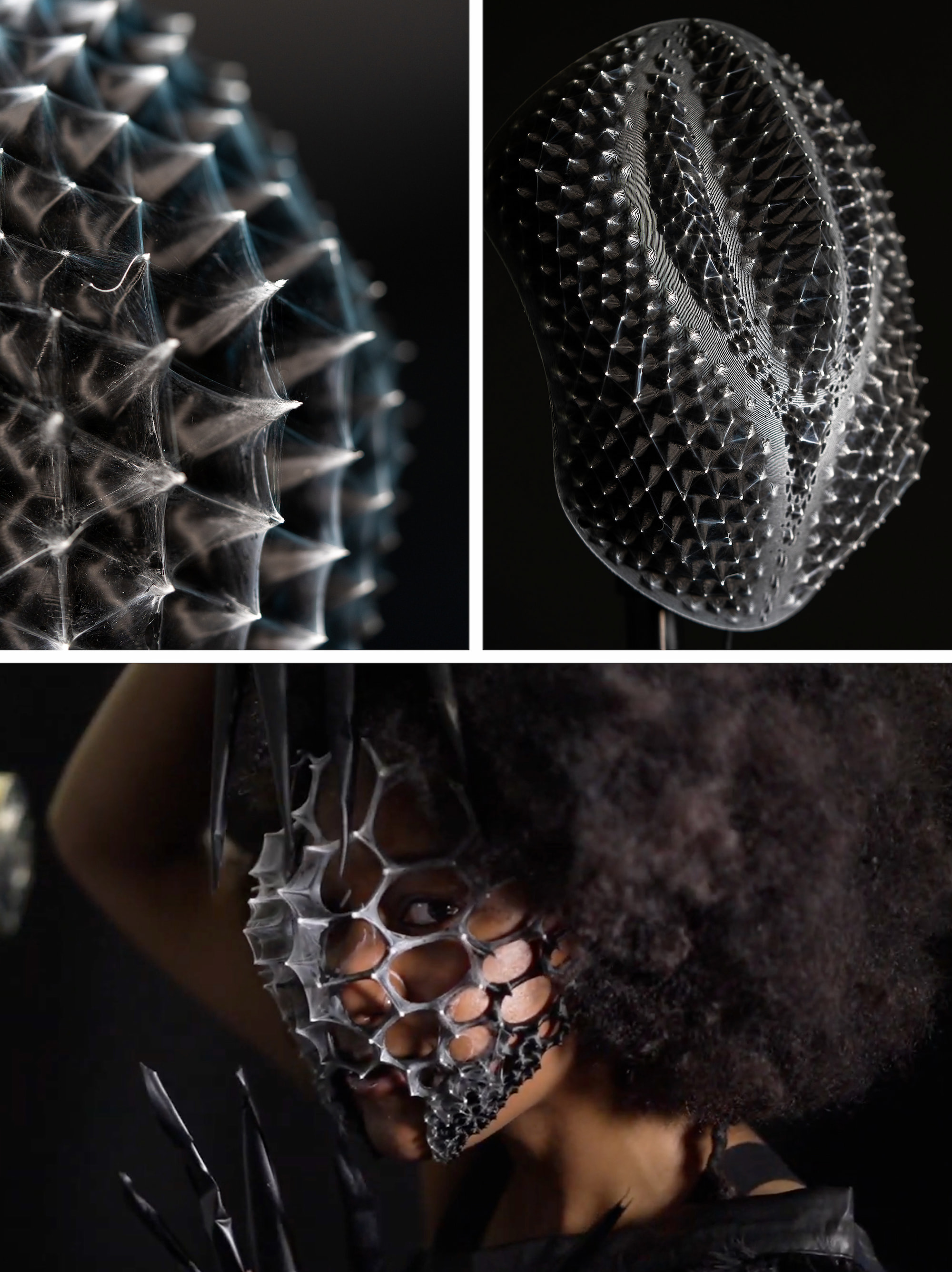}
  \caption{Electrospun keratin on conductive PLA organic mask prototypes}
  \label{fig:8_Mask}
\end{figure}

This mask (Figure~\ref{fig:8_Mask}) is electrospun from keratin extracted from hair donated by members of the research team. Hair is at once the body's most abundant waste product and one of its most culturally charged materials---a signifier of identity, heritage, and belonging. Dissolving and re-spinning hair from multiple individuals into a single nano-fiber membrane collapses separate identities into a shared material layer worn over the face. The wearer's own hair and nails---unprocessed biological keratin---remain visible in the frame, staging a confrontation between the body's native structuring of this protein and its re-structuring through invisible electrostatic force.

In fabrication, the piece synthesizes hierarchical network behaviors from the deposition studies above. The substrate employs a parametric hexagonal grid adapted to facial curvature, featuring perpendicular spikes at intersections. Acting as electric field intensifiers, these spikes guide nanofibers to bridge nodal gaps. By algorithmically modulating spike height, we created a dynamic gradient: taller spikes capture dense fibers for structural opacity, while lower profiles encourage translucent membranes. The fabrication of this non-planar geometry necessitated the full kinematic dexterity of the UR20 robotic platform. To overcome field shielding inherent to concave topologies, we developed a custom toolpath that continuously reorients the end-effector to track surface normals, ensuring stable deposition on steep slopes.

\subsection{\textit{Chrysalis}: Electrospun PVA Sculptural Garment on Wire}

\begin{figure} [h]
  \includegraphics[width=0.85\textwidth]{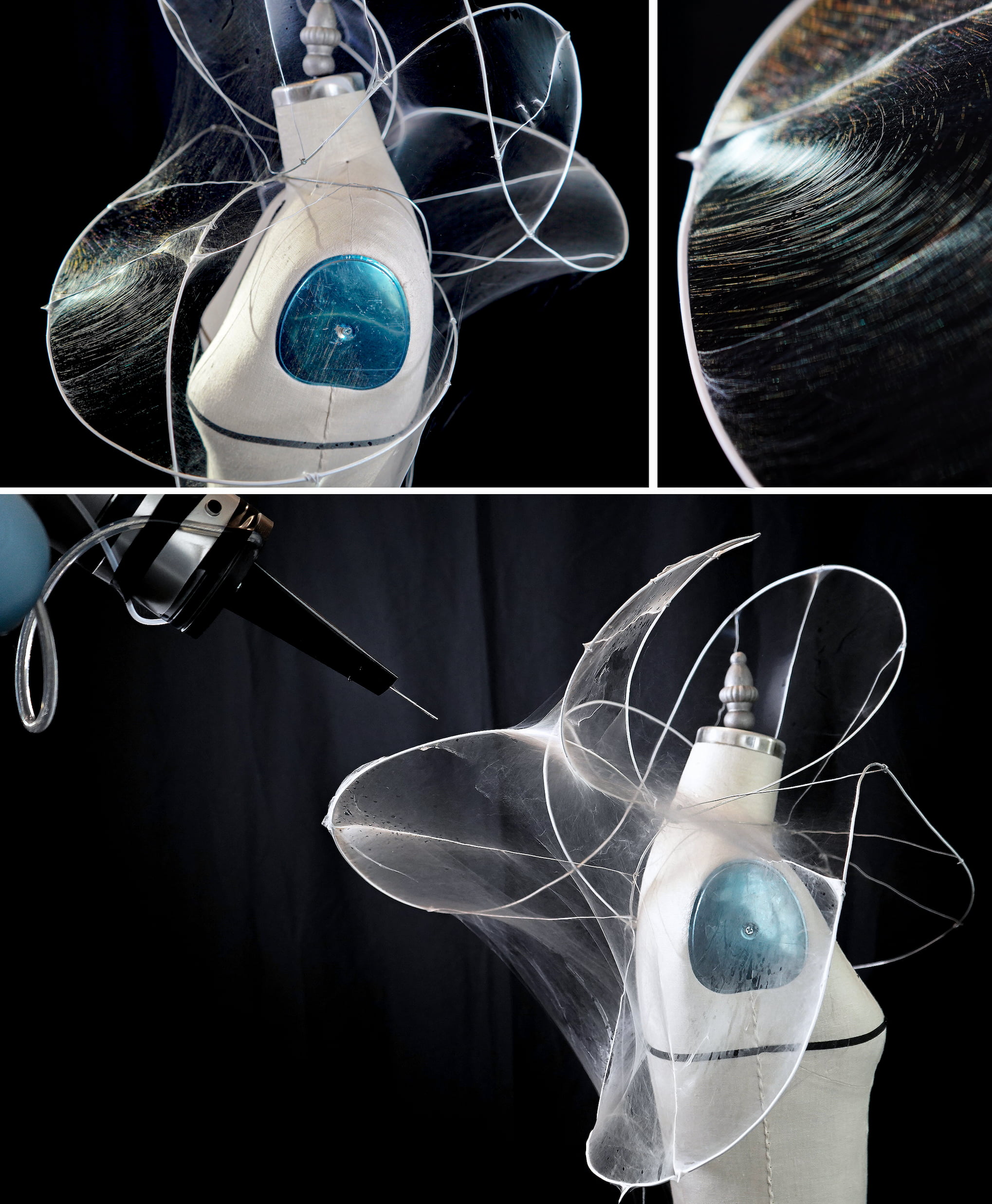}
  \caption{Electrospun PVA sculptural garment on wire scaffold using robotic electrospinner}
  \label{fig:9_Garment}
\end{figure}

Drawing on the morphology of Lepidopteran wings---where minimal venation supports vast, structurally efficient membranes---\textit{Chrysalis} (Figure~\ref{fig:9_Garment}) explores electrospinning as a metamorphic process. A liquid polymer solution is dissolved into a jet and re-constituted as a fibrous membrane whose geometry is found, not prescribed. The resulting garment records a negotiation between wire armature and field-driven material logic: dense structural ridges trace high-field paths along conductors, while translucent sub-membranes fill the interstices with emergent minimal-surface geometries visible at close range. Worn on the body, the piece transforms the wearer into both scaffold and environment---a participant in the field conditions that generate form.

In fabrication, the piece is built on a hand-sculpted wire chassis whose cantilevered loops challenge the material to bridge open topologies. Robotic orbital toolpaths intensify field gradients between spars; continued deposition forces fibers to span gaps up to 45\,cm, decoupling fabrication scale from 3D-printed scaffold constraints and generating emergent non-planar ruled surfaces between non-adjacent boundary curves.


\subsection{\textit{Invisible Ink}: Programmable Deposition Matrix}

\begin{figure} [h]
  \includegraphics[width=\textwidth]{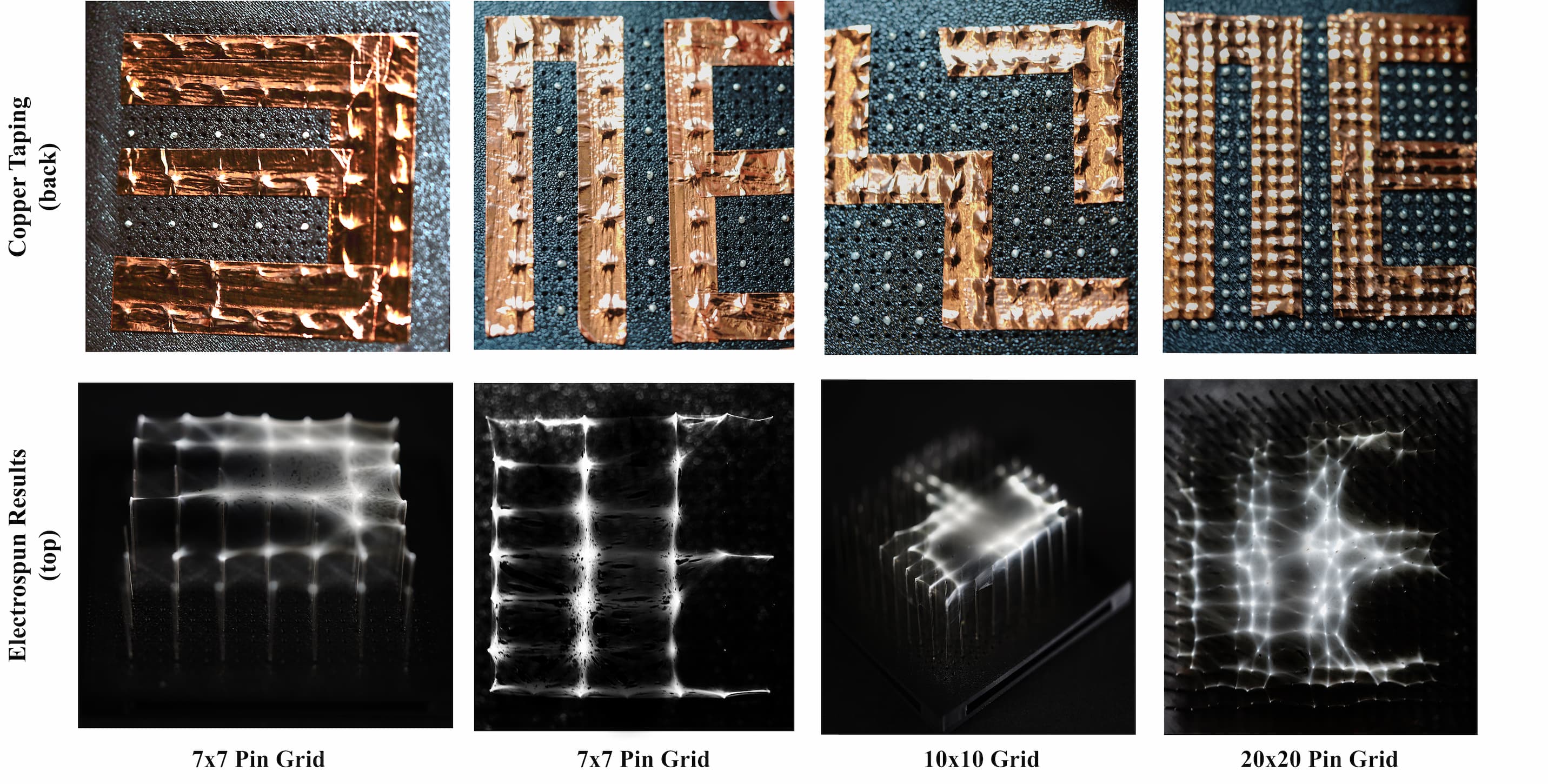}
  \caption{Programmable Deposition Matrix. (Top) Copper taping grounding pin array sections from the backside and (Bottom) resulting electrospun material deposition on front}
  \label{fig:10_Programmable}
\end{figure}

Where the preceding artifacts demonstrate how scaffold geometry conditions material form, \textit{Invisible Ink} (Figure~\ref{fig:10_Programmable}) inverts that relationship: the material image is authored not by physical geometry but by the selective activation of an invisible electric field. Fibers render the field pattern into visible matter---a material photograph of an unseen force, and a distillation of the core premise of \textit{Electrospun Fields}: invisible forces, made legible through material accumulation.

In this experimental setup, we utilized a modular pin array where specific nodes could be selectively grounded via manual circuitry. By grounding only selected coordinates, we effectively ``activate'' specific electrical field attractors while leaving adjacent pins inert. Consequently, electrospun fibers accumulated strictly along the energized pathways, allowing us to ``draw'' variable geometric patterns on a static physical tool. This demonstrates that the electric field, and thus the resulting material image, can be decoupled from permanent physical geometry, suggesting a future where dynamically actuated fields could function as a reconfigurable, low-resolution material display.

\section{Discussion \& Conclusion}

This research establishes \textit{Electrospun Fields} as a material--computational design method that shifts fabrication from geometric prescription to field-conditioned form-finding, treating electric fields as generative design conditions rather than constraints. The method enables ultra-light 3D membranes difficult to produce with planar collectors or geometry-driven deposition alone, while supporting biomaterials such as hair-derived keratin. As the demonstrative artifacts illustrate, these capacities open expressive territory---from the cultural resonance of body-derived keratin restructured as wearable membrane, to the layered form-finding of field-grown spatial textiles---positioning electrospinning as a medium for computational art.

The demonstrative artifacts show that behaviors such as cross-void suspension and hyperbolic self-organization can be reliably elicited within a scaffold taxonomy and programmable grounding strategy, supported by a reproducible robotic workflow that others can adopt and extend. A step-by-step tutorial and open-source repository documenting the platform assembly, material formulations, and scaffold designs are publicly available at \url{https://github.com/Critical-Matter-MIT-Media-Lab/Custom-3D-Electrospinning}. At the same time, outcomes remain partially non-deterministic and sensitive to material and environmental conditions; rather than a limitation, this defines a force-driven design space that privileges emergence and material intelligence.

Looking forward, bio-compatible keratin and silk suggest pathways for in-situ electrospinning onto the human body. As skin is conductive, it can serve as the grounded collector, enabling customized second skins such as medical dressings or performative textiles. Future work will integrate real-time 3D scanning and computer vision to adapt deposition to physiological movement, and extend programmable grounding toward reconfigurable, field-driven material displays.

Ultimately, \textit{Electrospun Fields} reimagines fabrication as a process of designing “with” material systems rather than designing “for” them. By treating natural forces—here, electric fields—as computational agents, the method shifts design toward learning from how form emerges, stabilizes, and grows under constraint. Fabrication becomes an inquiry into formation itself, where modeling operates at the level of conditions and interactions rather than predetermined outcomes. In this sense, \textit{Electrospun Fields} proposes a design language computed through natural forces, rendered in fiber, and form-found across space through material self-organization. 

\begin{acks}
We would like to acknowledge Frank (Haotian) Cong and Berfin Ataman for their invaluable contributions to the early development of the system. Our thanks also go to James Xiao for developing the syringe pump, and to Paolo Salvagione for his guidance on mechanical and electrical engineering. This research was developed in collaboration with the Rutledge Research Group at the MIT Department of Chemical Engineering; we extend our sincere gratitude to Nathan Ewell and Greg Rutledge for their expertise and generous assistance. Finally, we gratefully acknowledge the MIT--LUMA Grant for their generous financial support.
\end{acks}

\bibliographystyle{ACM-Reference-Format}
\bibliography{References}

\end{document}